\documentclass[10pt,journal]{IEEEtran}

\ifCLASSOPTIONcompsoc
  \usepackage[nocompress]{cite}
\else
  \usepackage{cite}
\fi

\usepackage{times}
\usepackage{epsfig}
\usepackage{graphicx}
\usepackage{amsmath}
\usepackage{amssymb}
\usepackage{multirow}
\usepackage{pifont}
\usepackage{booktabs}
\usepackage{epsfig}
\usepackage{booktabs,multirow}
\usepackage{graphics}
\usepackage{threeparttable}
\usepackage{color}
\usepackage[normalem]{ulem}
\usepackage{multirow}
\usepackage{float}
\usepackage{amsfonts}
\usepackage{bm}
\usepackage{bbm}
\usepackage{subfig}
\usepackage{enumitem}
\usepackage[numbers]{natbib}
 \usepackage{url}
\usepackage{array}
\usepackage[table]{xcolor}
\usepackage{colortbl}
\definecolor{bblue}{rgb}{0,150,230}
\definecolor{mygray}{gray}{.9}
\definecolor{myy}{RGB}{126,95,0}
\usepackage{bm}
\newcolumntype{I}{!{\vrule width 1pt}}

\definecolor{ggray}{RGB}{127,127,127}

\newcommand{\eg}[1]{\textit{e.g.,}}
\newcommand{\ie}[1]{\textit{i.e.,}}
\newcommand{\etc}[1]{\textit{etc}}

\makeatletter
\newcommand{\thickhline}{%
	\noalign {\ifnum 0=`}\fi \hrule height 1pt
	\futurelet \reserved@a \@xhline
}
\makeatother

\newcommand{\figref}[1]{Fig.\!~\ref{#1}}

\usepackage[pagebackref=true,breaklinks=true,letterpaper=true,colorlinks,bookmarks=false]{hyperref}

\usepackage{hyperref}

\usepackage{cleveref}
\crefname{section}{}{§§}
\Crefname{section}{}{§§}

\usepackage{caption}
\begin{document}

\title{DONet: Dual-Octave Network for Fast MR Image Reconstruction}

\author{Chun-Mei Feng, Zhanyuan Yang, Huazhu Fu, \IEEEmembership{Senior Member, IEEE}, Yong Xu, \IEEEmembership{Senior Member, IEEE},\\
Jian Yang, \IEEEmembership{Member, IEEE} and Ling Shao, \IEEEmembership{Fellow, IEEE}

\thanks{C.-M. Feng and Y. Xu are with the Shenzhen Key Laboratory of Visual Object Detection and Recognition, Harbin Institute of Technology (Shenzhen), 518055, China.~(Email: strawberry.feng0304@gmail.com; yongxu@ymail.com).~(Corresponding author: \textit{Yong Xu})}
\thanks{Z. Yang is with the School of Automation Engineering, University of Electronic Science and Technology of China, 611731, China. (Email: hawkyang0826@gmail.com)}
\thanks{H. Fu and L. Shao are with the Inception Institute of Artificial Intelligence, Abu Dhabi, UAE. (Email: hzfu@ieee.org; ling.shao@ieee.org)}

\thanks{J.~Yang is with the PCA Laboratory, Key Laboratory of Intelligent Perception and Systems for High-Dimensional Information of Ministry of Education, Nanjiang University of Science and Technology, Nanjiang 210094, China, and also with the Jiangsu Key Laboratory of Image and Video Understanding for Social Security, School of Computer Science and Engineering, Nanjing University of Science and Technology, Nanjing 210094, China (e-mail: csjyang@njust.edu.cn).}

\thanks{This work was done during the internship of Chun-Mei Feng at Inception Institute of Artificial Intelligence.}

}

\markboth{IEEE Transactions on Neural Networks and learning Systems}%
{Shell \MakeLowercase{\textit{et al.}}: Bare Demo of IEEEtran.cls for Computer Society Journals}

\IEEEtitleabstractindextext{%
\begin{abstract}

Magnetic resonance (MR) image acquisition is an inherently prolonged process, whose acceleration has long been the subject of research. 
This is commonly achieved by obtaining multiple undersampled images, simultaneously, through parallel imaging.
In this paper, we propose the Dual-Octave Network (DONet), which is capable of learning multi-scale spatial-frequency features from both the real and imaginary components of MR data, for parallel fast MR image reconstruction. 
More specifically, our DONet consists of a series of Dual-Octave convolutions (Dual-OctConv), which are connected in a dense manner for better reuse of features. In each Dual-OctConv, the input feature maps and convolutional kernels are first split into two components (\ie, real and imaginary), and then divided into four groups according to their spatial frequencies. 
Then, our Dual-OctConv conducts intra-group information updating and inter-group information exchange to aggregate the contextual information across different groups. 
Our framework provides three appealing benefits: (i) It encourages information interaction and fusion between the real and imaginary components at various spatial frequencies to achieve richer representational capacity. (ii) The dense connections between the real and imaginary groups in each Dual-OctConv make the propagation of features more efficient by feature reuse. (iii) DONet enlarges the receptive field by learning multiple spatial-frequency features of both the real and imaginary components.
Extensive experiments on two popular datasets (\ie, clinical knee and fastMRI), under different undersampling patterns and acceleration factors, demonstrate the superiority of our model in accelerated parallel MR image reconstruction.

\end{abstract}

}

\maketitle
\IEEEdisplaynontitleabstractindextext
\IEEEpeerreviewmaketitle

\section{Introduction}

\label{sec:introduction}
\IEEEPARstart{M}{agnetic} resonance (MR) imaging has become increasingly popular in radiology and medicine over the past decade, thanks to its advantages in being non-radiative, having a high spatial-resolution, and providing superior soft tissue contrast~\cite{1}. However, a major limitation of MR imaging is that it requires a much longer acquisition time than other imaging techniques, \eg, computed tomography (CT), X-Ray, and ultrasound~\cite{2,feng2021multi}. In addition, it is impossible to keep patients in the scanner for a long time and obtain clean data without motion artifacts. Therefore, accelerating MR reconstruction has become an urgent research problem, since it can in turn greatly accelerate MR imaging. Recently, significant efforts have been devoted to this task, which is typically achieved by reconstructing the desired full images from undersampled measured data~\cite{aggarwal2018modl,feng2021brain}.


\begin{figure}[t]
\centering
  \includegraphics[width=0.5\textwidth]{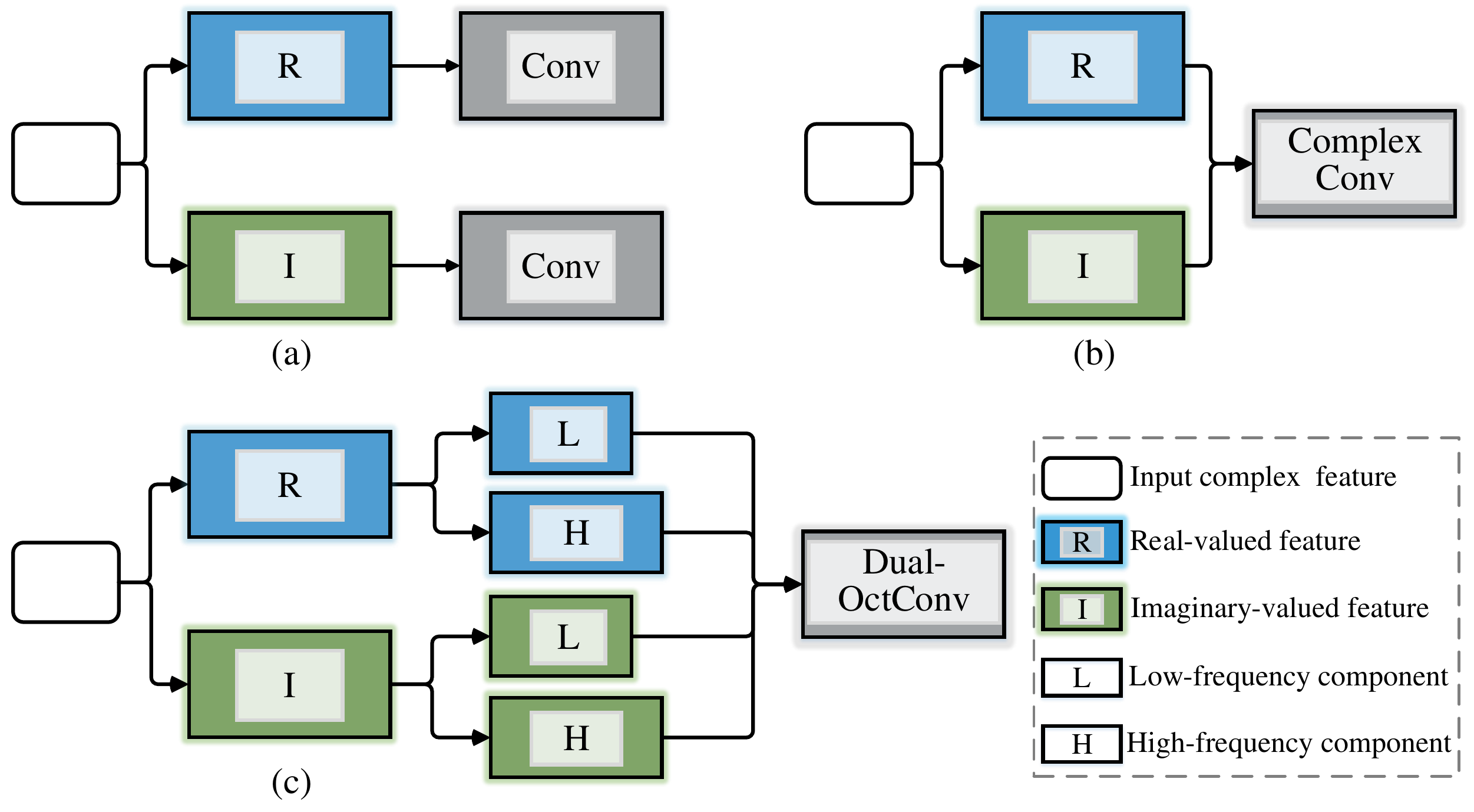}
  \caption{Motivation of our work. Previous methods utilize vanilla convolutions (a) to process the real- and imaginary-valued parts of an MR image independently, or complex convolutions (b) to jointly deal with the two parts. In contrast, we propose Dual-OctConv, which is a generalization of complex convolutions, to process complex-valued inputs in a multi-frequency space for more effective feature representations.}
  \label{movitation}
\end{figure}

Compressed Sensing (CS) has achieved significant progress in fast MR imaging, because sub-Nyquist sampling can significantly reduce the acquisition time by skipping some phase information. Specifically, CS-based methods overcome aliasing artifacts caused by the violation of the Shannon-Nyquist sampling theorem, by introducing additional prior knowledge of the images. CS-based methods employ sparse coding in the transformed domain (\eg~ undersampled $k$-space data) to naturally compress the MR image through a discrete cosine transform (DCT)~\cite{lingala2013blind,wang2014undersampled}, discrete Fourier transform (DFT)~\cite{hot2015compressed,gilbert2014recent}, discrete wavelet transform (DWT)~\cite{lai2016image,qu2012undersampled,chaari2011wavelet}, or dictionary learning~\cite{liu2020highly}. However, these traditional methods only exploit prior information of the images to-be-reconstructed, or involve very few reference images. For a large number of medical images acquired in clinical practice, traditional CS-based methods do not explore the potential regularities within them. Because CS-based methods require iterative optimization to find the optimal value, even if the CS involves only a small number of reference images, they still suffer from heavy computational overhead.

Recently, parallel MR imaging has been considered as one of the most important achievements in accelerated MR imaging~\cite{knoll2019deep,wang2017learning,chang2011kernel,chen2013calibrationless}. Multi-coil data for parallel imaging is composed of multiple physical receivers simultaneously recorded from different perspectives. Parallel MR imaging reconstructs the data points using the coil sensitivity profiles across multiple channels, from a small amount of $k$-space data. At present, parallel imaging is the default option for many scanning protocols. Most studies (\eg~ SENSE~\cite{pruessmann1999sense}, GRAPPA~\cite{griswold2002generalized,weller2013sparsity}, SPIRiT~\cite{lustig2010spirit}) take advantage of spatial sensitivity and gradient coding to reduce the amount of data required for reconstruction, thereby shortening the imaging time. However, a higher acceleration rate may introduce aliased artifacts and substantially reduce the signal-to-noise ratio (SNR) in clinical practice.


More recently, with the renaissance of deep neural networks, deep learning techniques, especially convolutional neural networks (CNNs)~\cite{zhou2021contrast,zhou2020matnet}, have been widely used for parallel MR imaging~\cite{ramani2010parallel,haldar2016p}. Since models are trained offline over large-scale data, only a few extra online samples are required for reconstruction. The model-based unrolling methods~\cite{12,chen2019model} combine mathematical structures (\eg~ variational inference, compressed sensing) with deep learning for fast MR image reconstruction. Moreover, extensive approaches~\cite{kwon2017parallel,schlemper2019sigma,schlemper2019data1,sriram2020grappanet,2} introduce end-to-end learnable models to remove the aliasing artifacts from images that are reconstructed from undersampled multi-coil $k$-space data. The mapping between a zero-filled $k$-space and fully-sampled MR image is automatically learned by CNNs, requiring no sub-problem division.


Most of the above approaches directly borrow vanilla convolutions used in standard CNNs for $k$-space data in MR image reconstruction. However, vanilla convolutions are designed for real-valued natural images, and cannot deal with complex-valued inputs. To solve this, early studies~\cite{wang} simply discarded the imaginary part or processed the real and imaginary parts independently for real-valued convolutions (see \figref{movitation}(a)). To avoid information loss, the complex convolution~\cite{Trabelsi2018DeepCN} was recently proposed to process complex-valued inputs and encourage information exchange between real and imaginary values (see \figref{movitation}(b)). Though impressive, existing complex convolution operations ignore the intrinsic multi-frequency property of MR images, leading to limited single-scale contextual information and high spatial redundancy in the final representations.

To address these limitations, we take a further step towards exploring multi-frequency representation learning in parallel MR image reconstruction (see~\figref{movitation}(c)). Specifically, we propose a novel Dual-Octave Network (DONet) which contains a series of Dual-Octave Convolutions (Dual-OctConv), enabling our model to learn multi-frequency representations of multi-coil MR images~\cite{chen2019drop}. The dense connections in the Dual-OctConv enable the feature propagation of the high- and low-frequency components of both the real and imaginary data. Unlike complex convolutions, our DONet processes the real (or imaginary) part of MR image features by factorizing it into high- and low-frequency components. The low-frequency component shares information across neighboring locations, and can thus be efficiently processed in low-resolution to enlarge the receptive field and reduce the spatial redundancy.
Finally, we combine the features of the real and imaginary parts for reconstruction. Benefiting from the Dual-OctConv, our network has a more powerful capability in multi-scale representation learning, and can thus better capture soft tissues (\eg~blood vessels, muscles) with varying sizes and shapes. 



Our main contributions are three-fold: \textbf{First}, we propose a novel DONet containing multiple Dual-OctConvs connected in a dense  manner for multi-frequency feature reuse, and demonstrate their ability to capture multi-scale contextual information. \textbf{Second}, we employ the Dual-OctConv to deal with complex-valued inputs in a multi-frequency representation space, and encourage information exchange and fusion across various frequency domains. The Dual-OctConv is a generalization of the standard complex convolution, and endows our model several appealing characteristics (\eg~a larger receptive field, higher flexibility, and more computational efficiency). 
\textbf{Third}, our model shows significant performance improvements against state-of-the-art algorithms on a clinical knee dataset and the fastMRI dataset.

Compared with our preliminary conference version~\cite{feng2021dual}, this paper provides the following extensions. (1) We generalize the architecture of the Dual-Octave (corresponding to DONet$^\dagger$ in this paper) from~\cite{feng2021dual} to a more powerful version (\ie, DONet). This is done by employing dense connections for better feature reuse. Our DONet is demonstrated to achieve significantly better performance than Dual-Octave~\cite{feng2021dual} (\S\ref{sec:ml}). (2) We further apply the proposed DONet to the currently largest raw MRI dataset, fastMRI, to demonstrate the advantages of our method over  other deep learning models (\S\ref{sec:ex}). (3) Finally, we also provide a more inclusive and insightful overview of recent work on fast parallel MRI reconstruction (\S\ref{sec:re}).

\begin{figure*}[t]
\centering
  \includegraphics[width=\textwidth]{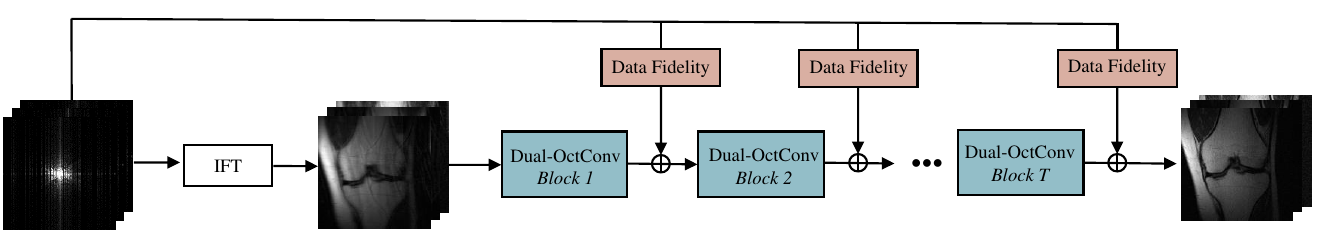}
  \caption{Architecture of our DONet for parallel MR image reconstruction. The input is a set of zero-filled multi-coil $k$-space measurements, while the output is the reconstructed multi-channel MR image. $\rm IFT$ represents the 2D inverse Fourier transform. See \S\ref{sec:framework} for details.}
  \label{flowchart}
\end{figure*}

\section{Related Work}
\label{sec:re}
\subsection{Deep Learning in MR Image Reconstruction}
Ever since the pioneering works introducing CNNs for computer vision tasks, such as image classification and face recognition, researchers have made substantial efforts to improve medical and clinical practice using deep learning techniques~\cite{feng2021MINet,feng2021T2Net}. Wang \textit{et al.}~\cite{wang} proposed the first deep learning based MR image reconstruction framework, which learns the mapping between fully-sampled single-coil MR images and their counterpart data reconstructed from a zero-filled undersampled $k$-space. A large number of networks have since been developed for MR image reconstruction, especially non-parallel reconstruction~\cite{sun2019deep}. For example, \cite{4} proposed a model-based unrolling method, which formulates the algorithm within a deep neural network, and trained the network with a small amount of data. As an end-to-end method, ~\cite{han2018deep} employed U-Net to model a domain adaptation structure that removes aliasing artifacts from corrupted images. Zhu \textit{et al.}~\cite{zhu2018image} proposed the automated transform by manifold approximation (AUTOMAP) framework, which they implemented using a deep neural network to learn a  mapping between the sensor and image domain. Han \textit{et al.}~\cite{han2018deep} proposed a deep learning approach with domain adaptation which uses a large amount of X‐ray computed tomography (CT) or synthesized radial MR data and is fine‐tuned with a small amount of real radial MR data. Dongwook \textit{et al.}~\cite{lee2018deep} trained amplitude and phase, respectively, using a residual network to reduce the computational cost of the reconstruction algorithm. With the pioneering works on Generative Adversarial Networks (GANs) for natural image synthesis, GAN-based methods have also made great progress in MR image reconstruction~\cite{quan2018compressed, yang2017dagan, 15}. For example, Tran \textit{et al.}~\cite{quan2018compressed} designed an adversarial model for CS-MR image reconstruction, accurately interpolating and restoring data in an undersampled $k$-space data, and with a loss of data consistency. Moreover, image refinement on specific network structures with different training objectives, such as the adversarial loss~\cite{yang2017dagan} or perceptual loss~\cite{quan2018compressed}, has gained more attention.

In parallel imaging, one representative model is the variational network (VN-Net)~\cite{12}, which combines the mathematical structure of the variational model with deep learning for fast multi-coil MR image reconstruction. Another model-based deep framework~\cite{chen2019model} was designed with a split Bregman iterative algorithm to achieve accurate reconstruction from multi-coil undersampled $k$-space data. To obtain high-fidelity reconstructions, GrappaNet~\cite{sriram2020grappanet} was proposed to combine traditional parallel imaging methods with deep neural networks. Recently, complex-valued representations have demonstrated superiority in processing complex-valued inputs~\cite{Trabelsi2018DeepCN}. For example,~\cite{2} applied complex convolutions to jointly process real and imaginary values for comprehensive feature representations. In contrast, our approach represents complex-valued input features in a multi-frequency space. The Dual-OctConv, proposed for processing such multi-frequency data, can capture richer contextual knowledge, leading to significant improvement in performance.

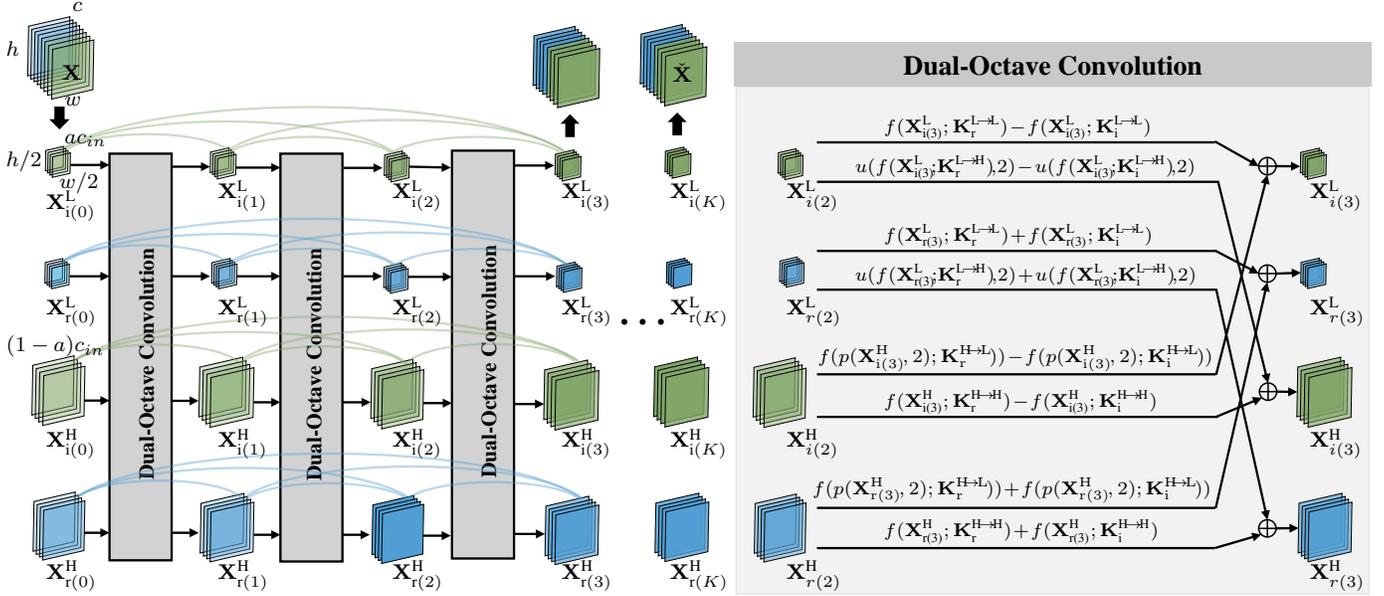
\begin{figure*}[h]
\centering
  \includegraphics[width=1\textwidth]{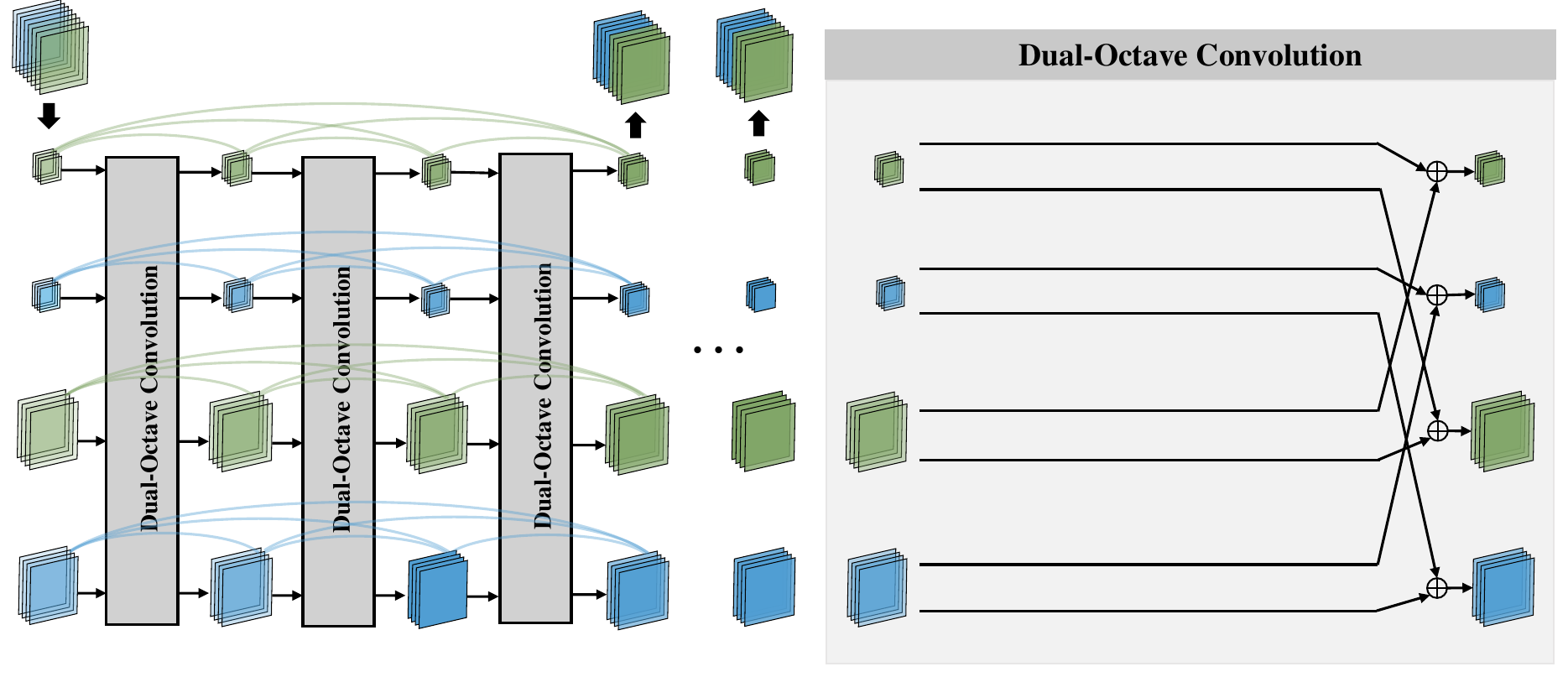}
  \put(-520,208){\footnotesize $h$}
  \put(-498,189){\footnotesize $w$}
  \put(-495,225){\footnotesize $c$}
  \put(-520,166){\footnotesize $h/2$}
  \put(-500,158){\footnotesize $w/2$}
  \put(-498,175){\footnotesize $ac_{in}$}
  \put(-520,96){\footnotesize $(1-a)c_{in}$}
  \put(-499,199){\footnotesize $\mathbf{X}$}
  \put(-268,198){\footnotesize $\check{\mathbf{X}}$}
  \put(-505,150){\footnotesize $\mathbf{X}_{\text{i}(0)}^{\text{L}}$}
  \put(-505,110){\footnotesize $\mathbf{X}_{\text{r}(0)}^{\text{L}}$}
  \put(-505,60){\footnotesize $\mathbf{X}_{\text{i}(0)}^{\text{H}}$}
  \put(-505,10){\footnotesize $\mathbf{X}_{\text{r}(0)}^{\text{H}}$}
  \put(-440,154){\footnotesize $\mathbf{X}_{\text{i}(1)}^{\text{L}}$}
  \put(-440,110){\footnotesize $\mathbf{X}_{\text{r}(1)}^{\text{L}}$}
  \put(-440,60){\footnotesize $\mathbf{X}_{\text{i}(1)}^{\text{H}}$}
  \put(-440,10){\footnotesize $\mathbf{X}_{\text{r}(1)}^{\text{H}}$}
  \put(-374,154){\footnotesize $\mathbf{X}_{\text{i}(2)}^{\text{L}}$}
  \put(-374,110){\footnotesize $\mathbf{X}_{\text{r}(2)}^{\text{L}}$}
  \put(-374,60){\footnotesize $\mathbf{X}_{\text{i}(2)}^{\text{H}}$}
  \put(-374,10){\footnotesize $\mathbf{X}_{\text{r}(2)}^{\text{H}}$}
  \put(-310,154){\footnotesize $\mathbf{X}_{\text{i}(3)}^{\text{L}}$}
  \put(-310,110){\footnotesize $\mathbf{X}_{\text{r}(3)}^{\text{L}}$}
  \put(-310,60){\footnotesize $\mathbf{X}_{\text{i}(3)}^{\text{H}}$}
  \put(-310,10){\footnotesize $\mathbf{X}_{\text{r}(3)}^{\text{H}}$}
  \put(-269,154){\footnotesize $\mathbf{X}_{\text{i}(K)}^{\text{L}}$}
  \put(-269,110){\footnotesize $\mathbf{X}_{\text{r}(K)}^{\text{L}}$}
  \put(-269,60){\footnotesize $\mathbf{X}_{\text{i}(K)}^{\text{H}}$}
  \put(-269,10){\footnotesize $\mathbf{X}_{\text{r}(K)}^{\text{H}}$}
  \put(-225,154){\footnotesize $\mathbf{X}_{i(2)}^{\text{L}}$}
  \put(-225,110){\footnotesize $\mathbf{X}_{r(2)}^{\text{L}}$}
  \put(-225,60){\footnotesize $\mathbf{X}_{i(2)}^{\text{H}}$}
  \put(-225,10){\footnotesize $\mathbf{X}_{r(2)}^{\text{H}}$}
  \put(-27,154){\footnotesize $\mathbf{X}_{i(3)}^{\text{L}}$}
  \put(-27,110){\footnotesize $\mathbf{X}_{r(3)}^{\text{L}}$}
  \put(-27,60){\footnotesize $\mathbf{X}_{i(3)}^{\text{H}}$}
  \put(-27,10){\footnotesize $\mathbf{X}_{r(3)}^{\text{H}}$}
  \put(-188,179){\scriptsize $f(\mathbf{X}_{\text{i(3)}}^{\text{L}};\mathbf{K}_{\text{r}}^{\text{L}\!\rightarrow\!\text{L}})\!-\!f(\mathbf{X}_{\text{i(3)}}^{\text{L}};\mathbf{K}_{\text{i}}^{\text{L}\!\rightarrow\!\text{L}})$}
  \put(-199,164){\scriptsize $u(f(\mathbf{X}_{\text{i(3)}}^{\text{L}}\!;\!\mathbf{K}_{\text{r}}^{\text{L}\!\rightarrow\!\text{H}})\!,\!2)\!-\!u(f(\mathbf{X}_{\text{i(3)}}^{\text{L}}\!;\!\mathbf{K}_{\text{i}}^{\text{L}\!\rightarrow\!\text{H}})\!,\!2)$}
  \put(-188,138){\scriptsize $f(\mathbf{X}_{\text{r(3)}}^{\text{L}};\mathbf{K}_{\text{r}}^{\text{L}\!\rightarrow\!\text{L}})\!+\!f(\mathbf{X}_{\text{r(3)}}^{\text{L}};\mathbf{K}_{\text{i}}^{\text{L}\!\rightarrow\!\text{L}})$}
  \put(-199,123){\scriptsize $u(f(\mathbf{X}_{\text{r(3)}}^{\text{L}}\!;\!\mathbf{K}_{\text{r}}^{\text{L}\!\rightarrow\!\text{H}})\!,\!2)\!+\!u(f(\mathbf{X}_{\text{r(3)}}^{\text{L}}\!;\!\mathbf{K}_{\text{i}}^{\text{L}\!\rightarrow\!\text{H}})\!,\!2)$}
  \put(-213,92){\scriptsize 
  $f(p(\mathbf X_{\rm i(3)\!}^{\rm H},2);\mathbf K_{\rm r}^{{\rm H\!}\!\rightarrow\!{\rm L\!}}))\!-\!f(p(\mathbf X_{\rm i(3)\!}^{\rm H\!},2);\mathbf K_{\rm i\!}^{{\rm H\!}\!\rightarrow\!{\rm L\!}}))$}
  \put(-188,75){\scriptsize $f(\mathbf{X}_{\text{i(3)}}^{\text{H}};\mathbf{K}_{\text{r}}^{\text{H}\!\rightarrow\!\text{H}})\!-\!f(\mathbf{X}_{\text{i(3)}}^{\text{H}};\mathbf{K}_{\text{i}}^{\text{H}\!\rightarrow\!\text{H}})$}
  \put(-215,42){\scriptsize   
  $f(p(\mathbf X_{\rm r(3)\!}^{\rm H},2);\mathbf K_{\rm r}^{{\rm H\!}\!\rightarrow\!{\rm L\!}}))\!+\!f(p(\mathbf X_{\rm r(3)\!}^{\rm H\!},2);\mathbf K_{\rm i}^{{\rm H\!}\!\rightarrow\!{\rm L\!}}))$}
  \put(-188,26){\scriptsize $f(\mathbf{X}_{\text{r(3)}}^{\text{H}};\mathbf{K}_{\text{r}}^{\text{H}\!\rightarrow\!\text{H}})\!+\!f(\mathbf{X}_{\text{r(3)}}^{\text{H}};\mathbf{K}_{\text{i}}^{\text{H}\!\rightarrow\!\text{H}})$}

  \caption{Detailed design of our Dual-OctConv block. $\mathbf{X}\!\in\!\mathbb{C}^{c\!\times\!h\!\times\!w}$ represents the input complex-valued feature maps, and $\mathbf{Y}\!\in\!\mathbb{C}^{c\!\times\!h\!\times\!w}$ indicates the corresponding output feature maps, modulated by the Dual-OctConv. $u$ and $p$ denote the upsampling and average pooling operations, respectively. Please see Eq.~\eqref{eq:5} for more details.}
  \label{dense}
\end{figure*}

\subsection{Multi-Scale Representation Learning}
Multi-scale information has proven effective in various computer vision tasks (\eg, image classification, object detection, semantic segmentation). Especially in the era of deep learning, multi-scale representation has been successfully applied in various fields due to its strong robustness and generalization ability. Several strategies have been proposed for this, yielding significant performance improvement in a number of tasks. For example,~\cite{ke2017multigrid} proposed a multi-grid network to propagate and integrate information across multiple scales for image classification. Multi-scale information has also been proven effective in restoring image details for image enhancement~\cite{nah2017deep,ren2016single,li2018multi}. Seungjun \textit{et al.}~\cite{nah2017deep} proposed a multi-scale convolutional neural network together with a multi-scale loss function for dynamic scene deblurring. In addition, various well-known techniques (\eg, FPN~\cite{lin2017feature} and PSP~\cite{zhao2017pyramid}) have been proposed for learning multi-scale representations in object detection and segmentation tasks. For example, Lin \textit{et al.}~\cite{lin2017feature} proposed a top-down feature pyramid architecture with lateral connections at all scales for object detection. Recently, the Octave convolution~\cite{chen2019drop} was proposed to learn multi-scale features based on the spatial frequency model~\cite{campbell1968application,devalois1990spatial}, greatly improving performance in natural image and video recognition. 

In this work, we demonstrate the appealing properties of the Octave convolution for accelerated parallel MR image reconstruction, which helps to capture multi-scale information from features of multiple spatial frequencies. Based on this, we propose a novel Dual-OctConv for accelerated parallel MR image reconstruction, which enables our model to capture details of vasculatures and tissues with varying sizes and shapes, yielding high-fidelity reconstructions.

\section{Methodology}
\label{sec:ml}
\subsection{Problem Formulation}
MR scanners acquire $k$-space data through the receiver coils and then utilize an inverse multidimensional Fourier transform to obtain the final MR images.
In parallel imaging, multiple receiver coils are used to simultaneously acquire $k$-space data from the target under scanning.


Let $\mathbf A\!=\!\mathbf M\!\mathbf F\!\in\!\mathbb{C}^{M\times N}$ denote the undersampled Fourier encoding matrix, where $\mathbf F$ is the multidimensional Fourier transform, and $\mathbf M$ is an undersampled mask operator.
In parallel imaging, the same mask is used for all coils.
The undersampled $k$-space data from each coil can be expressed as:
\begin{equation}
\mathbf y_i = \mathbf A (\mathbf{S}_i\mathbf x), \label{XX}
\end{equation}
where $i=1,2,...,c$, with $c$ denoting the number of coils,
$\mathbf x\!\in\!\mathbb{C}^{N\times 1}$ is the ground truth MR image,
$\mathbf y_i\!\in\!\mathbb{C}^{M\times 1}(M\!<\!<\!N)$ is the undersampled $k$-space data for the $i$-th coil,
and $\mathbf{S}_i$ is a complex-valued diagonal matrix encoding the sensitivity map of the $i$-th coil.
The coil sensitivities, which are measured by each coil, modulate the $k$-space data.
The coil configuration and interactions with the anatomical structures under scanning can affect the coil sensitivities, so $\mathbf{S}_i$ changes across different scans.
In addition, the obtained image will contain aliasing artifacts if the inverse Fourier transform is directly applied to the undersampled $k$-space data. For single-coil MR imaging, $i=1$.

We can reconstruct $\hat{\mathbf x}$ with prior knowledge of its properties, which can be formulated as the following optimization problem:
\begin{equation}
	\hat{\mathbf x}=\mathop{\arg \min}_{\mathbf x}\sum_{i=1}^c \  \| {\mathbf y_i - \mathbf A (\mathbf{S}_i\mathbf x) } \|^2_2+ \ \lambda \Psi(\mathbf x),
	\label{eq:object_function}
\end{equation}
where $\Psi$ is a regularization function and $\lambda$ controls the trade-off between the two terms.

The problem presented in Eq.~\eqref{eq:object_function} can be effectively resolved using CNNs, which avoids time-consuming numerical optimization and the need of a coil sensitivity map.
During training, we update the network weights by minimizing an $\ell_1$ loss function, 
\begin{equation}
	\hat{\boldsymbol{\theta}}= \mathop{\arg \min}_{\boldsymbol{\theta}} \frac{1}{N}\sum_{n=1}^{N} \| {\mathbf{x'}(n) - f_{\boldsymbol{\theta}}( \mathbf{y'}(n) )} \|_1,
\label{eq:theta}
\end{equation}
where $\mathbf{y'}(n)$ is the $n$-th multi-channel image obtained from the zero-filled $k$-space data, $\mathbf{x'}(n)$ is the $n$-th multi-channel ground truth image, $N$ is the total number of training samples, and $f_{\boldsymbol{\theta}}(\cdot)$ is an end-to-end mapping function parameterized by $\boldsymbol{\theta}$, which contains a large number of adjustable network weights.
Training with Eq.~\eqref{eq:theta} can reconstruct the expected MR images, but the original information of the data acquired in the $k$-space cannot be well preserved.
If we incorporate the undersampled $k$-space data into the data fidelity at the training stage, the network can yield improved reconstruction results.
For this purpose, we add the data fidelity units in our network, as in \cite{2}.
After the network is trained, we obtain a set of optimal parameters $\boldsymbol{\hat \theta}$ for the reconstruction of a multi-channel image, and predict this image via
$\mathbf{\hat{x}'} = f_{\hat{\boldsymbol{\theta}}}(\mathbf{y'})$.
Finally, we use an adaptive coil combination method~\cite{2} to obtain the expected MR image from $\mathbf{\hat{x}'}$.

\begin{figure}[t]
  \centering
  \includegraphics[width=0.5\textwidth]{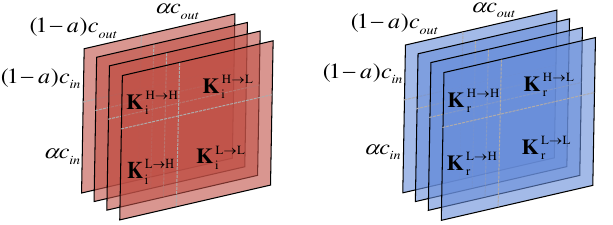}
  \caption{The Dual-OctConv kernels. Red and blue squares correspond to the imaginary and real kernels, respectively.}
  \label{(b)}
\end{figure}

\subsection{Framework Overview}\label{sec:framework}

~\figref{flowchart} illustrates the architecture of our DONet, which simultaneously considers both $k$-space data fidelity and image space proximity to achieve high-fidelity reconstruction. Specifically, given an undersampled multi-coil $k$-space measurement as input, we first convert it to aliased multi-channel images via an inverse Fourier transform (IFT). The transformed image is then fed into a cascade of $T$ Dual-OctConv blocks, with each block followed by a data fidelity unit~\cite{2}. The data fidelity unit helps to preserve the original $k$-space information during training. Formally, the $t$-th data fidelity unit can be written as:
\begin{equation}\label{eq:dc}
      \mathbf{X}^{t} =\mathbf{F}^{-1}((\mathbf{1}-\mathbf{M}) \mathbf{F} (\check{\mathbf{X}}^t)+\mathbf{M} \mathbf{y}),
\end{equation}
where $\check{\mathbf{X}}^t$ denotes the reconstructed output image of the $t$-th Dual-OctConv block, while $\mathbf{X}^t$ represents the output after applying the $t$-th data fidelity unit, which will be fed into the $t\!+\! 1$-th Dual-OctConv block as input. $\mathbf{F}^{-1}$ denotes the inverse Fourier transform.

Next, we will describe the details of our Dual-OctConv block in \S\ref{sec:block} and the Dual-OctConv layer in \S\ref{sec:conv}.

\subsection{Dual-Octave Convolutional Block}\label{sec:block}
 We present the detailed architecture of our Dual-OctConv block, which is employed to learn comprehensive, multi-scale feature representations for reconstruction. In the following paragraphs, the superscript `$t$' is omitted for conciseness, unless necessary.

 As shown in~\figref{dense}, each Dual-OctConv block consists of $K$ Dual-OctConv layers $\{\mathcal{D}_k\}_{k=1}^K$. These layers accecpt $\mathbf{X}$ as input and produce the reconstruction $\check{\mathbf{X}}$, in a sequential manner. More specifically, to obtain rich multi-scale context information, we first represent the multi-channel input with complex filters, and then decompose it into low and high spatial frequency parts. Let $\mathbf X \!\in\!\mathbb{C}^{c\times h\times w}=\mathbf X_{\rm r (0)}+i\mathbf X_{\rm i (0)}$ be the input complex feature maps with $c$, $h$, and $w$ denoting the number of channels, height, and width, respectively. The $\mathbf X_{\rm r(0)}$ and $\mathbf X_{\rm i(0)}$ are the real and imaginary parts of the complex data, respectively. As illustrated in~\figref{dense}, we initially split the complex input feature maps into low- and high-frequency groups: $\mathbf{X}_{\rm r(0)} \!=\! \{ \mathbf X_{\text{r}(0)}^{\rm L}, \mathbf X_{\text{r}(0)}^{\rm H} \}$, $\mathbf X_{\rm{i}(0)} \!=\! \{ {\mathbf X_{\text{i}(0)}^{\rm L}, \mathbf X_{\text{i}(0)}^{\rm H}} \}$, where $\mathbf X^{\rm H}\!\in\!\mathbb{C} ^{(1-\alpha) c \times h \times w}$ captures the high-frequency fine details of the data and $\mathbf X^{\rm L}\!\in\!\mathbb{C} ^{\alpha c \times 0.5h \times 0.5w}$ determines the low-frequency image contrast. Here, $\alpha\!\in\![0,1]$ controls the ratio of channels that are allocated to low-frequency and high-frequency feature maps. Note that the Dual-OctConv becomes the standard complex convolution~\cite{Trabelsi2018DeepCN} when $\alpha\!=\!0$. Then, for the $k$-th Dual-OctConv layer $\mathcal{D}_k$, we have:
\begin{equation}
    \begin{aligned}
   \mathbf X_{\text{r}(k)}^{\text L}, & \mathbf X_{\text{r}(k)}^{\text H}, \mathbf X_{\text{i}(k)}^{\rm L}, \mathbf X_{\text{i}(k)}^{\rm H} \\
   & = \mathcal{D}_k ( \mathbf X_{\text{r}(k-1)}^{\rm L}, \mathbf X_{\text{r}(k-1)}^{\rm H}, \mathbf X_{\text{i}(k-1)}^{\rm L}, \mathbf X_{\text{i}(k-1)}^{\rm H}),
   \end{aligned}
   \label{Dk}
\end{equation}%
where $\mathbf X_{\text{r}(k-1)}^{\rm L}$, $\mathbf X_{\text{r}(k-1)}^{\rm H}$, $\mathbf X_{\text{i}(k-1)}^{\rm L}$, $\mathbf X_{\text{i}(k-1)}^{\rm H}$ are the inputs, and $\mathbf X_{\text{r}(k)}^{\rm L}$, $\mathbf X_{\text{r}(k)}^{\rm H}$, $\mathbf X_{\text{i}(k)}^{\rm L}$, $\mathbf X_{\text{i}(k)}^{\rm H}$ are the outputs of $\mathcal{D}_k$. Then, for each block, we obtain the final feature representations at the $K$-th Dual-OctConv layer, which are combined to obtain the reconstruction:
\begin{equation}
\check{\mathbf{X}} = u(c(\mathbf X_{\text{r}(K)}^{\rm L}, \mathbf X_{\text{i}(K)}^{\text L}), 2)+c(\mathbf X_{\text{r}(K)}^{\text{H}}, \mathbf X_{\text{i}(K)}^{\rm H}),
\end{equation}
where $u(\cdot,z)$ denotes the upsampling operation with a factor of $z$. Here, we use the nearest neighbor interpolation. $c(\cdot, \cdot)$ denotes the concatenation operation. The output $\check{\mathbf{X}}$ is then fed into the data fidelity layer Eq.~(\ref{eq:dc}) as well as the subsequent Dual-OctConv block for cascade reconstruction.

In addition, we emphasize that the sequential pipeline mentioned above conducts individual feature learning layer-by-layer, neglecting the feature fusion among different layers, which often produces more comprehensive feature representations~\cite{szegedy2015going,huang2017densely}. To this end, we improve the sequential Dual-OctConv block by introducing dense connections in an interleaved manner. Formally, for the output of the $k$-th ($k\!\geq\!1$) Dual-OctConv layer, its features are modulated as follows:
\begin{equation}
\begin{split}
    \mathbf X_{\text{r}(k)}^{\rm L} &= \mathcal{C}_k([\mathbf X_{\text{r}(k)}^{\rm L} , \mathbf X_{\text{r}(k-1)}^{\rm L}, \cdots,\mathbf X_{\text{r}(0)}^{\rm L}]),\\
    \mathbf X_{\text{i}(k)}^{\rm L} &= \mathcal{C}_k([\mathbf X_{\text{i}(k)}^{\rm L} , \mathbf X_{\text{i}(k-1)}^{\rm L}, \cdots, \mathbf X_{\text{i}(0)}^{\rm L}]),\\
    \mathbf X_{\text{r}(k)}^{\rm H} &= \mathcal{C}_k([\mathbf X_{\text{r}(k)}^{\rm H} , \mathbf X_{\text{r}(k-1)}^{\rm H}, \cdots, \mathbf X_{\text{r}(0)}^{\rm H}]),\\
    \mathbf X_{\text{i}(k)}^{\rm H} &= \mathcal{C}_k([\mathbf X_{\text{i}(k)}^{\rm H} , \mathbf X_{\text{i}(k-1)}^{\rm H}, \cdots, \mathbf X_{\text{i}(0)}^{\rm H}]).\\
\end{split}
\label{eq:7}
\end{equation}
Here, the feature maps at the $k$-th layer are firstly concatenated the features at all previous layers, and then processed with a function $\mathcal{C}_k$ to learn collective knowledge from all these feature maps. Following~\cite{huang2017densely}, we implement $\mathcal{C}_k$ as: BN-ReLU-Conv($1\!\times\!1$)-BN-ReLU-Conv($3\!\times\!3$). 
Such dense connections enable our model to reuse features from previous layers with high computational efficiency, since there is no need to relearn redundant feature maps. They also help to alleviate the gradient vanishing problem when the number of layers in our network increases. In our experiments, we demonstrate that the reconstruction performance significantly improves after introducing these dense connections.

\subsection{Dual-Octave Convolutional Layer}\label{sec:conv}

Now we present the proposed Dual-Octave convolutional layer, which fuses the real and imaginary parts of a complex valued input. For the $k$-th layer $\mathcal{D}_k$, we denote $\mathbf{X}_{(k-1)}\!=\!\mathbf{X}_{\text{r}(k-1)} + i\mathbf{X}_{\text{i}(k-1)}$ as its input. We convolve $\mathbf{X}_{(k-1)}$ with a complex filter matrix $\mathbf K_{(k)} = \mathbf K_{\text{r}(k)}+i\mathbf K_{\text{i}(k)}$ as follows:
\begin{equation}
\!\!\left[\!\begin{array}{ccc}
\Re(\mathbf K_{(k)}* \mathbf X_{(k-1)})\\
\Im(\mathbf K_{(k)}* \mathbf X_{(k-1))}
\end{array} 
\!\right] \!=\! {\left[ \!\begin{array}{ccc}
\mathbf K_{\text{r}(k)} \!\!&\!\! -\mathbf K_{\text{i}(k)}\\
\mathbf K_{\text{i}(k)} \!\!&\!\!   \mathbf K_{\text{r}(k)}
\end{array} 
\!\right]} \!*\!
{
\left[\!\begin{array}{ccc}
\mathbf X_{\text{r}(k-1)} \\
\mathbf X_{\text{i}(k-1)} 
\end{array} 
\!\right]},
\end{equation}
where the matrices $\mathbf K_{\text{r}(k)}$ and $\mathbf K_{\text{i}(k)}$ represent real and imaginary kernels, respectively. Note that all the kernels and feature maps are expressed by real matrices since the complex arithmetics are simulated by real-valued entities.

\begin{table*}[t]
 \centering
 \caption{Quantitative comparison of state-of-the-art methods under different undersampling patterns on the clinical knee dataset. Best and second best results are marked in red and blue, respectively.}
 \resizebox{\textwidth}{!}{
 \setlength\tabcolsep{2pt}
 \renewcommand\arraystretch{1.3}
 \begin{tabular}{r||cc|cc|cc|cc|cc|cc|cc|cc}
  \thickhline\rowcolor{mygray}
  {undersampling pattern} & \multicolumn{4}{c|}{1D Uniform}  &  \multicolumn{4}{c|}{1D Cartesian} & \multicolumn{4}{c|}{2D Random} & \multicolumn{4}{c}{2D Radial} \\ \cline{2-17} 
  \rowcolor{mygray}
  {acceleration rate} & \multicolumn{2}{c|}{3x} & \multicolumn{2}{c|}{5x}  & \multicolumn{2}{c|}{3x} & \multicolumn{2}{c|}{5x} & \multicolumn{2}{c|}{3x} & \multicolumn{2}{c|}{5x} & \multicolumn{2}{c|}{4x} & \multicolumn{2}{c}{6x} \\ \cline{2-17}
  \rowcolor{mygray}
       & PSNR & SSIM & PSNR & SSIM & PSNR & SSIM & PSNR & SSIM & PSNR & SSIM & PSNR & SSIM & PSNR & SSIM & PSNR & SSIM \\ \hline\hline
  Zero-filing & 24.406 & 0.676  & 23.579 & 0.655 & 25.922 & 0.726  & 24.550 & 0.685  & 30.540 & 0.827  & 27.078 & 0.750  & 31.026 & 0.826 &28.107 & 0.766 \\
  SPIRiT &29.385 & 0.700  &28.300 & 0.676   &32.310 & 0.801  &31.222 & 0.782 &32.179 & 0.786  &32.258 & 0.812   &30.308 &0.720  &29.061 &0.702 \\
  L1-SPIRiT &29.815 & 0.847  &27.353 & 0.788   &33.346 & 0.887  &30.912 & 0.837 &38.597 & 0.937  &34.071 & 0.887   &37.004  &0.919 &34.149 &0.881\\
  VN-Net &35.436 & 0.907  &32.730 & 0.858   &36.364 & 0.912  &33.236 & 0.866 &38.409 & {\color{red}0.956}  &35.734 & 0.923   & 37.956 & 0.930 &34.609  &0.907 \\
  ComplexMRI &34.989 & 0.909  &32.803 & 0.873   &35.957 & 0.916  &34.126 & 0.876 &39.563 & 0.946  &37.315 & 0.908   &38.098  &0.933 &35.768 &0.904\\ \hline
  DONet$^\dagger$ &{\color{blue}36.243} & {\color{red}0.919}  &{\color{blue}34.128} & {\color{blue}0.885}   &{\color{blue}37.029} & {\color{blue}0.923}  &{\color{blue}34.944} & {\color{blue}0.884} &{\color{blue}39.964} & 0.948  &{\color{blue}38.279} & {\color{red}0.930}   &{\color{red}38.607}  &{\color{red}0.935} &{\color{red}36.584} &{\color{blue}0.908} \\ 
   DONet &\textcolor[rgb]{1.00,0.00,0.00}{36.338} & {\color{blue}0.918}  &\color{red}{34.526} & \textcolor[rgb]{1.00,0.00,0.00}{0.892}   &\color{red}{37.074} & \textcolor[rgb]{1.00,0.00,0.00}{0.931}  &\textcolor[rgb]{1.00,0.00,0.00}{35.400} & \textcolor[rgb]{1.00,0.00,0.00}{0.902} &\textcolor[rgb]{1.00,0.00,0.00}{40.652} & {\color{blue}0.953}  &\textcolor[rgb]{1.00,0.00,0.00}{38.428} & {\color{red}0.930}   &{\color{blue}38.387}  &{\color{red}0.935} &{\color{blue}36.281} &{\color{red}0.909} \\ \hline
  
 \end{tabular}}
 \label{t1}
\end{table*}

As shown in~\figref{(b)}, the complex filter matrix is further expressed as $\mathbf K_{\rm r}^{\rm H} = [\mathbf K_{\text{r}(k)}^{{\rm H}\rightarrow {\rm L}},\mathbf K_{\text{r}(k)}^{{\rm H}\rightarrow {\rm H}} ]$, $\mathbf K_{\text{i}(k)}^{\rm H} \!=\! [\mathbf K_{\text{i}(k)} ^{{\rm H}\rightarrow {\rm L}},\mathbf K_{\text{i}(k)}^{{\rm H}\rightarrow {\rm H}} ]$, $\mathbf K_{\text{r}(k)}^{\rm L} \!=\! [\mathbf K_{\text{r}(k)}^{{\rm L}\rightarrow {\rm H}},\mathbf K_{\text{r}(k)}^{{\rm L}\rightarrow {\rm L}} ]$, $\mathbf K_{\text{i}(k)}^{\rm L} \!=\! [\mathbf K_{\text{i}(k)}^{{\rm L}\rightarrow {\rm H}},\mathbf K_{\text{i}(k)}^{{\rm L}\rightarrow {\rm L}} ]$ to convolve with $\mathbf X_{\text{r}(k-1)}^{\rm L}$, $\mathbf X_{\text{i}(k-1)}^{\rm L}$, $\mathbf X_{\text{r}(k-1)}^{\rm H}$ and $\mathbf X_{\text{i}(k-1)}^{\rm H}$. Then, $\mathcal{D}_k$ can be formulated as:
\begin{equation}
\begin{split}
     \!\mathbf X_{\text{r}(k)}^{\rm L} \!=\!&f(\mathbf X_{\text{r}(k-1)}^{\rm L};\mathbf K_{\text{r}(k)}^{{\rm L}\rightarrow {\rm L}})+u(f(\mathbf X_{\text{r}(k-1)}^{\rm L};\mathbf K_{\text{r}(k)}^{{\rm L}\rightarrow H}),2) \\
     &\!\!\!\!\!+f(\mathbf X_{\text{r}(k-1)}^{\rm L};\mathbf K_{\text{i}(k)}^{{\rm L}\rightarrow {\rm L}})+u(f(\mathbf X_{\text{r}(k-1)}^{\rm L};\mathbf K_{\text{i}(k)}^{{\rm L}\rightarrow {\rm H}}),2),\\
     \!\mathbf X_{\text{i}(k)}^{\rm L} \!=\!& f(\mathbf X_{\text{i}(k-1)}^{\rm L};\mathbf K_{\text{r}(k)}^{{\rm L}\rightarrow {\rm L}})+u(f(\mathbf X_{\text{i}(k-1)}^{\rm L};\mathbf K_{\text{r}(k)}^{{\rm L}\rightarrow {\rm H}}),2)\\
     &\!\!\!\!\!-f(\mathbf X_{\text{i}(k-1)}^{\rm {\rm L}};\mathbf K_{\text{i}(k)}^{{\rm L}\rightarrow {\rm L}})-u(f(\mathbf X_{\text{i}(k-1)}^{\rm L};\mathbf K_{\text{i}(k)}^{{\rm L}\rightarrow {\rm H}}),2),\\
     \!\mathbf X_{\text{r}(k)}^{\rm H} \!=\!&f(\mathbf X_{\text{r}(k-1)}^{\rm H};\mathbf K_{\text{r}(k)}^{{\rm H}\rightarrow {\rm H}})+f(p(\mathbf X_{\text{r}(k-1)}^{\rm H},2);\mathbf K_{\text{r}(k)}^{{\rm H}\rightarrow {\rm L}}))\\
     &\!\!\!\!\!+f(\mathbf X_{\text{r}(k-1)}^{\rm H};\mathbf K_{\text{i}(k-1)}^{{\rm H}\rightarrow {\rm H}})+f(p(\mathbf X_{\text{r}(k-1)}^{\rm H},2);\mathbf K_{\text{i}(k)}^{{\rm H}\rightarrow {\rm L}})),\\
     \!\mathbf X_{\text{i}(k)}^{\rm H} \!=\!&f(\mathbf X_{\text{i}(k-1)}^{\rm H};\mathbf K_{\text{r}(k)}^{{\rm H}\rightarrow {\rm H}})+f(p(\mathbf X_{\text{i}(k-1)}^{\rm H},2);\mathbf K_{\text{r}(k)}^{{\rm H}\rightarrow {\rm L}}))\\
     &\!\!\!\!\!-f(\mathbf X_{\text{i}(k-1)}^{\rm H};\mathbf K_{\text{i}(k)}^{{\rm H}\rightarrow {\rm H}})-f(p(\mathbf X_{\text{i}(k-1)}^{\rm H},2);\mathbf K_{\text{i}(k)}^{{\rm H}\rightarrow {\rm L}})),\\
\end{split}
\label{eq:5}
\end{equation}
where $f(\mathbf X;\mathbf K)$ indicates the convolution with kernel $\mathbf K$, $u(\mathbf X,k)$ denotes the upsampling operation with a factor of $k$ via nearest interpolation, and $p(\mathbf X,z)$ represents an average pooling layer with kernel size $z\!\times\!z$.
The real and imaginary parts are fused with the operations $\left\{ {\rm L}\rightarrow {\rm L},{\rm H}\rightarrow {\rm H} \right\}$ and $\left\{ {\rm H}\rightarrow {\rm L},{\rm L}\rightarrow {\rm H} \right\}$, which correspond to the information updating and exchanging between high- and low-frequency feature maps.
Therefore, our Dual-OctConv is able to enlarge the receptive fields of the low-frequency feature maps both in the real and imaginary parts. To put this into perspective, after convolving the low-frequency feature maps of the real and imaginary parts ( $\mathbf X_{\text{r}(k-1)}^{\rm L}$, $\mathbf X_{\text{i}(k-1)}^{\rm L}$) with $z \times z$ complex convolutional kernels, the receptive fields of both  achieve a 2$\times$ enlargement compared to the vanilla convolution.
Thus, our Dual-OctConv has a strong ability to capture rich context information at different scales.

\section{Experiments}
\label{sec:ex}

\begin{table*}[t]
 \centering
 \caption{Quantitative comparison of state-of-the-art methods under different undersampling patterns on fastMRI dataset. Best and second best results are marked in red and blue, respectively. }
 \resizebox{\textwidth}{!}{
 \setlength\tabcolsep{2pt}
 \renewcommand\arraystretch{1.3}
 \begin{tabular}{r||cc|cc|cc|cc|cc|cc|cc|cc}
  \thickhline\rowcolor{mygray}
  {undersampling pattern} & \multicolumn{4}{c|}{1D Uniform}  &  \multicolumn{4}{c|}{1D Cartesian} & \multicolumn{4}{c|}{2D Random} & \multicolumn{4}{c}{2D Radial} \\ \cline{2-17}
  \rowcolor{mygray}
  {acceleration rate} & \multicolumn{2}{c|}{3x} & \multicolumn{2}{c|}{5x}  & \multicolumn{2}{c|}{3x} & \multicolumn{2}{c|}{5x} & \multicolumn{2}{c|}{3x} & \multicolumn{2}{c|}{5x} & \multicolumn{2}{c|}{3x} & \multicolumn{2}{c}{5x} \\ \cline{2-17}
  \rowcolor{mygray}
       & PSNR & SSIM & PSNR & SSIM & PSNR & SSIM & PSNR & SSIM & PSNR & SSIM & PSNR & SSIM & PSNR & SSIM & PSNR & SSIM \\ \hline\hline
  Zero-filing &22.981  &0.681   &22.600  &0.662  &22.553  &0.689   &22.274  &0.671   &29.292  &0.833   &25.614  &0.760   &31.308  &0.849  &28.460 &0.787  \\
  SPIRiT &27.305 &0.601   &27.185 &0.610    &29.387 &0.694   &28.133 &0.648  &30.950 &0.725   &30.293 &0.735    &29.778 &0.681  &28.099 &0.611 \\
  L1-SPIRiT &29.597 &0.827   &26.501 &0.774    &28.757 &0.835   &28.000 &0.801  &37.275 &0.911   &33.107 &0.852    &36.896  &0.907 &34.420 &0.868\\
  VN-Net &31.032 &0.860   &28.641 &0.810    &32.105 &0.875   &31.544 &0.853  &37.113 &0.907  &34.726 &0.813    &36.516  &0.894  &33.899  &0.825 \\
  ComplexMRI &31.548 & 0.867  &{\color{blue}29.922} &{\color{blue}0.829}  &32.773 & 0.888  &31.590 & 0.852 &37.751 & 0.926  &35.227 & 0.891   &37.238  &0.922 &34.753 &0.882\\ \hline
  DONet$^\dagger$ &{\color{blue}32.415} & {\color{blue}0.874}  &29.895 &{\color{blue}0.829}   &{\color{blue}33.204} & {\color{blue}0.891}  &{\color{blue}32.576} &{\color{blue}0.862} & {\color{blue}38.058} &{\color{blue}0.927}    &{\color{blue}35.911} & {\color{blue}0.894}   &{\color{blue}37.523}  &{\color{red}0.930} &{\color{blue}35.112} &{\color{blue}0.884} \\ 
   DONet &\textcolor[rgb]{1.00,0.00,0.00}{33.890} & {\color{red}0.888}  &\textcolor[rgb]{1.00,0.00,0.00}{32.780} & \textcolor[rgb]{1.00,0.00,0.00}{0.856}   &\textcolor[rgb]{1.00,0.00,0.00}{35.094} & \textcolor[rgb]{1.00,0.00,0.00}{0.904}  &\textcolor[rgb]{1.00,0.00,0.00}{33.644} & \textcolor[rgb]{1.00,0.00,0.00}{0.868} &\textcolor[rgb]{1.00,0.00,0.00}{38.316} & {\color{red}0.928}  &\textcolor[rgb]{1.00,0.00,0.00}{36.385} & {\color{red}0.896}   &{\color{red}37.905}  &{\color{blue}0.925} &{\color{red}35.413} &{\color{red} 0.886} \\ \hline
  
 \end{tabular}}
 \label{t2}
\end{table*}

\begin{figure*}[!htb]
\centering
  \includegraphics[width=\textwidth]{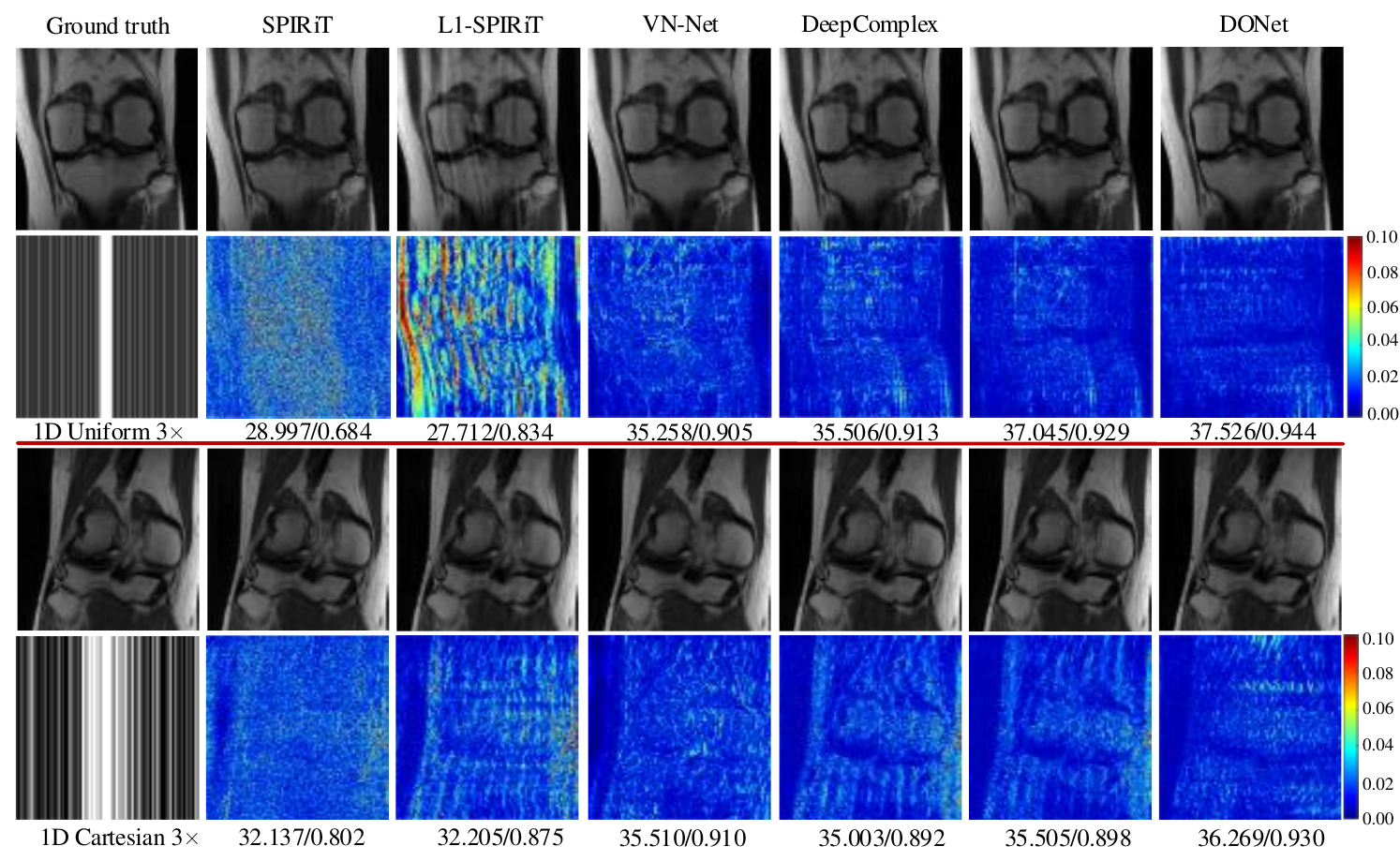}
  \put(-141,306){\small DONet$^\dagger$}
  \caption{Comparison of different methods in terms of reconstruction accuracy on the clinical knee dataset, with \textbf{1D undersampling patterns} and a \textbf{3$\times$} acceleration rate. Reconstruction results and error maps are presented with corresponding quantitative measurements in PSNR/SSIM.}
  \label{1D}
\end{figure*}

\begin{figure*}[!htb]
\centering
  \includegraphics[width=\textwidth]{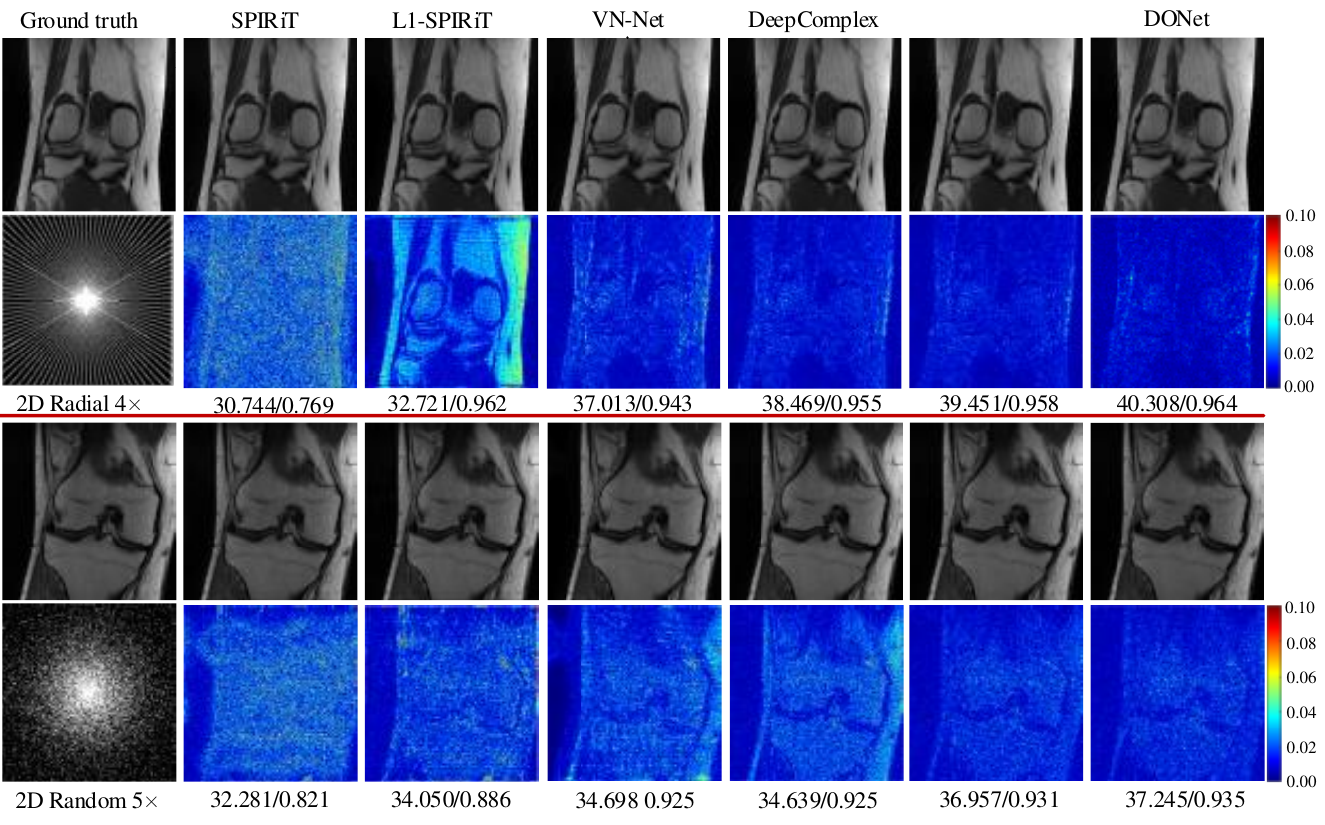}
  \put(-141,310){\small DONet$^\dagger$}
  \caption{Comparison of different methods in terms of reconstruction accuracy on the clinical knee dataset, with \textbf{2D undersampling patterns} and \textbf{4$\times$} and \textbf{5$\times$} acceleration rates. Reconstruction results and error maps are presented with corresponding quantitative measurements in PSNR/SSIM.}
  \label{2D}
\end{figure*}

\begin{figure}[t]
\centering
  \includegraphics[width=0.49\textwidth]{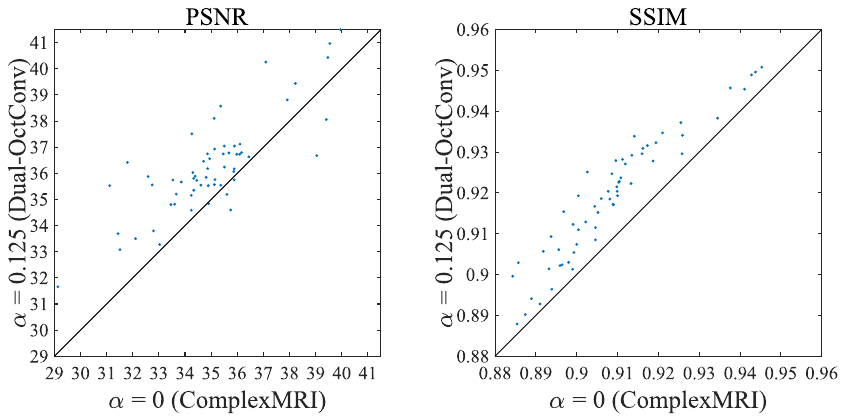}
  \caption{Quantitative comparison of Dual-OctConv and the baseline convolution ($\alpha\!=\!0$) in terms of PSNR and SSIM. Note that ratio $\alpha\!=\!0$ is equivalent to the complexMRI model.}
  \label{withoutoct}
\end{figure}

\begin{figure}[t]
\centering
  \includegraphics[width=0.49\textwidth]{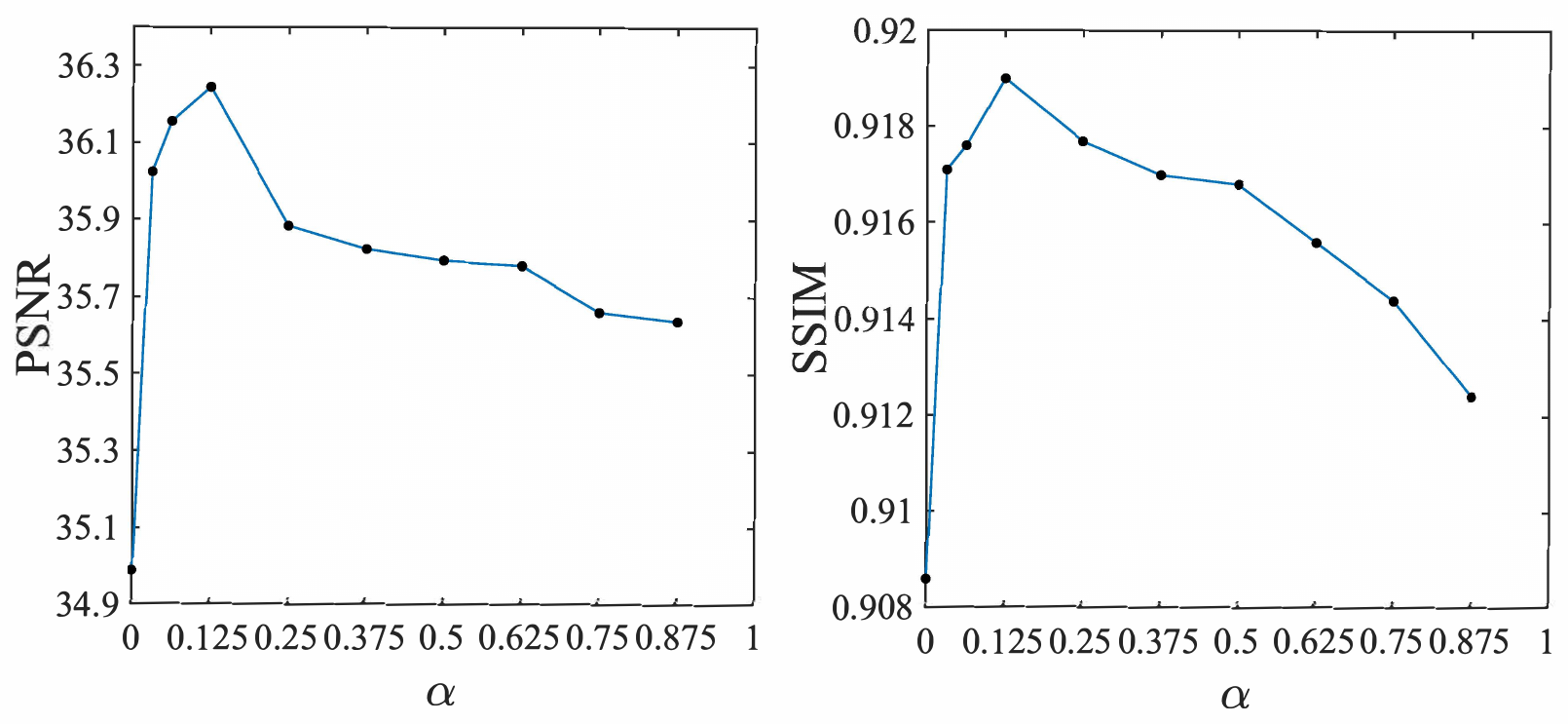}
  \caption{Analysis of spatial frequency ratio ($\alpha$) in terms of PSNR and SSIM.}
  \label{alpha}
\end{figure}

\begin{figure}[t]
	\centering
	\includegraphics[width=0.49\textwidth]{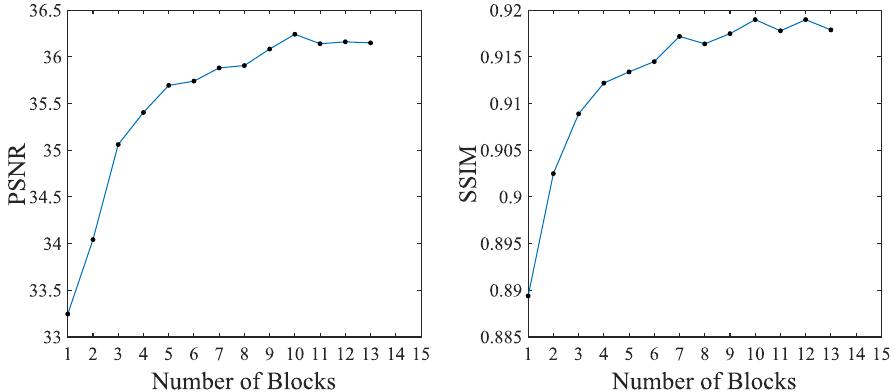}
	\caption{Performance comparison of our network with respect to the number of Dual-OctConv blocks.}
	\label{block}
\end{figure}

\begin{figure}[t]
\centering
  \includegraphics[width=0.49\textwidth]{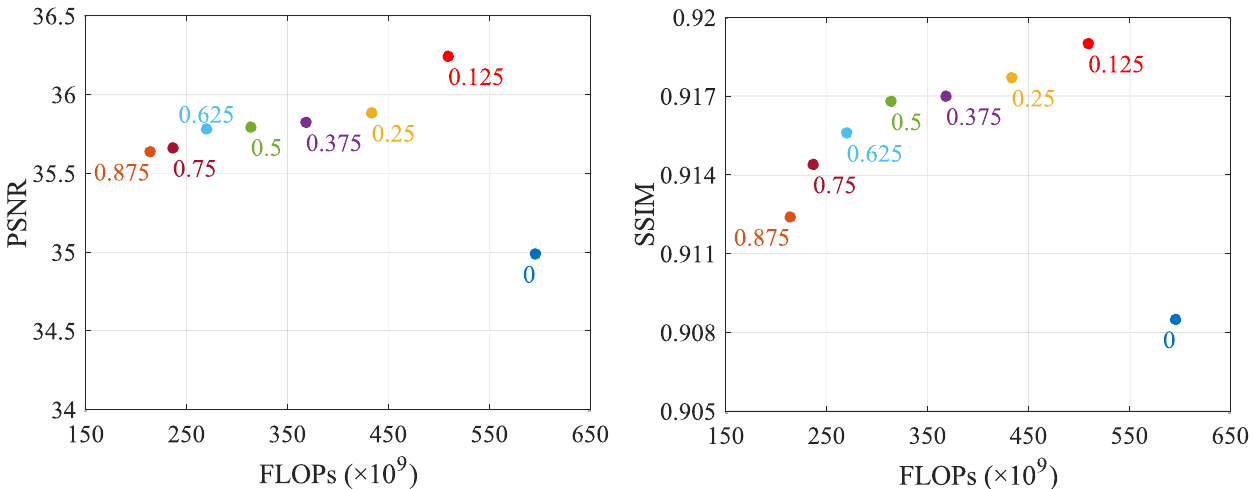}
  \caption{FLOPs analysis with respect to spatial frequency ratios ($\alpha$). The number below each point is the value of $\alpha$. We see that, under various settings ($0\!<\!\alpha\!<\!1$), our DONet is always more efficient and accurate than the baseline model ($\alpha\!=\!0$).}
  \label{flops}
\end{figure}

\subsection{Datasets}
We introduce a clinical knee~\cite{12} dataset and the fastMRI~\cite{zbontar2018fastmri} dataset under different undersampling patterns and acceleration factors to evaluate our method. The clinical multi-coil fully-sampled MR knee dataset is acquired using a clinical 3T Siemens Magnetom Skyra scanner with a sequence called ``Coronal Spin Density Weighted without Fat Suppression''. 
The imaging protocol is detailed as follows: 15-channel knee coil, matrix size 320$\times$320$\times$20, TR=2750 ms, TE=27 ms, and in-plane resolution = 0.49$\times$0.44 mm$^2$. There are 20 subjects in total with the following information: 5 female/15 male, age 15-76, and BMI 20.46-32.94. We randomly select fourteen patients for training, three for validation, and three for testing. For the fastMRI dataset, we randomly select 400 patients from the multi-coil knee dataset using a Proton Density (PD) weighted sequence for training, 40 for validation, and 60 for testing. In our experiments, we exclude the first few slices of each volume since the frontal slices are much noisier than the other slices, making the distribution of frontal slices different from the rest. More details of the fastMRI dataset can be obtained from~\cite{zbontar2018fastmri}. 

The pre-defined undersampling masks are used to obtain the undersampled measurements. In our experiments, we adopt four different $k$-space undersampling patterns, including 1D uniform, 1D Cartesian, 2D random, and 2D radial. Examples of the undersampling patterns are illustrated in~\figref{1D} and~\figref{2D}. For 1D uniform, 1D Cartesian, and 2D random masks, the acceleration rate is set to 3$\times$ and 5$\times$. For the 2D radial mask, 4$\times$ and 6$\times$ accelerations are adopted.

\subsection{Implementation Details}
We implement our model using Tensorflow 1.14 and perform experiments using an NVIDIA 1080Ti GPU with 11GB of memory.
Following~\cite{2}, we initialize the magnitude and phase of the complex parameters using Rayleigh and uniform distributions, respectively.
The network is trained using the Adam optimizer \cite{2} with initial learning rate 0.001 and weight decay 0.95.
The batch size is set to four and the convolutional kernel size is set to $3\times 3$. Each complex convolutional layer has 64 feature maps, except for the last layer, which is determined by the concatenated real and imaginary channels of the data. The spatial frequency ratio $\alpha$ is set to $0.125$ by default.

To demonstrate their effectiveness, we compare our DONet and DONet$^\dagger$ (without dense connections) with several state-of-the-art parallel MR imaging approaches, including traditional methods (SPIRiT \cite{lustig2010spirit} and L1-SPIRiT \cite{murphy2010clinically}) as well as CNN-based methods (VN-Net \cite{12} and ComplexMRI \cite{2}). All these methods are trained on the same dataset with their default settings. For CNN-based methods, we re-train them according to the specifications with TensorFlow, using their default parameter settings.

\subsection{Quantitative Evaluation }
We use peak signal-to-noise ratio (PSNR) and structural similarity index measure (SSIM)~\cite{2} for quantitative evaluation. Table~\ref{t1} reports the average PSNR and SSIM results with respect to different undersampling patterns and acceleration factors on the clinical dataset. As can be seen, DONet$^\dagger$ obtains consistent performance improvements against all baseline methods, demonstrating the superiority of our Dual-OctConv. Moreover, by incorporating dense connections, our full model (\ie, DONet) further improve the performance significantly across various settings.

Additionally, we observe that the undersampling patterns greatly affect the quality of reconstruction. For instance, the 2D sampling masks generally outperform the 1D masks. Another important observation is that the reconstruction becomes more difficult when the acceleration rate increases.



In particular, our model significantly outperforms previous methods under extremely challenging settings (\eg, 2D masks with 5$\times$ and 6$\times$ acceleration). This can be attributed to the powerful capability of our DONet in aggregating rich contextual information of real and imaginary data. Moreover, the overall results in Table~\ref{t1} show the strong robustness of our model under various undersampling patterns and acceleration rates. For example, our method restores more information and with minimum artifacts, as demonstrated in~\figref{1D} and ~\figref{2D} with error maps.

In Table~\ref{t2}, we provide reconstruction evaluations of our model on fastMRI, which is currently the largest MR image dataset. As we can see, DONet obtains the best reconstruction results among different undersampling patterns and acceleration rates. In particular, under 1D uniform undersampling patterns with 5$\times$ acceleration, our DONet improves the reconstruction PSNR from 29.992 dB to 32.780 dB and SSIM from 0.829 to 0.856, compared to the best previous state-of-the-art fusion method, ComplexMRI. Similarly, we observe that our DONet yields outstanding results on both the 1D uniform and 1D Cartesian undersampling patterns with 3$\times$ and 5$\times$ acceleration. More importantly, even without the dense connection, our DONet$^\dagger$ achieves compelling reconstruction results compared to the state-of-the-art methods. Under the 2D radial pattern with 3$\times$ acceleration, DONet$^\dagger$ achieves SSIM = 0.930. All the results demonstrate that our DONet preserves powerful and rich contextual information for real and imaginary data.


\subsection{Qualitative Evaluation }
For qualitative analysis, we first show the reconstructed images and corresponding error maps for 1D uniform and 1D Cartesian sampling with a 3$\times$ acceleration rate in~\figref{1D}. In general, 1D Cartesian masks provide better results in comparison to 1D uniform masks. Our method provides the best-quality reconstructed images and significantly reduces prediction errors. In contrast, the baseline methods yield large prediction errors and show unsatisfactory performance.

We next examine the results for 2D radial masks with a 4$\times$ acceleration rate and 2D random masks with a 5$\times$ acceleration rate. As shown in~\figref{2D}, CNN-based methods significantly improve the results with less errors and clearer structures, in comparison with SPIRiT and L1-SPIRiT. In particular, our DONet produces higher-quality images with clear details and minimum artifacts. The superior performance is owed to the fact that our method can effectively aggregate the information of various spatial frequencies present in the real and imaginary parts of an MR image.

\subsection{Ablation Studies}

The crucial part of our DONet is the Dual-OctConv. It is thus important to prove the effectiveness of the proposed convolution. First, we study the effects of Dual-OctConv. For comparison, we build a baseline convolution by setting $\alpha\!=\!0$, which turns our method into a standard complex convolution. We conduct experiments on the test set of $in$ $vivo$ dataset with 60 complex-valued images under the uniform undersampling mask with a 3$\times$ acceleration rate. As illustrated in~\figref{withoutoct}, our Dual-OctConv significantly outperforms the baseline model, especially in terms of SSIM. This reveals the superiority of the proposed method in improving the reconstruction.


Secondly, we investigate the influence of the spatial frequency ratio $\alpha$ for reconstruction. The ratio determines the receptive fields in both the real and imaginary parts, and also influences the fusion of these parts at multiple spatial frequencies. As shown in~\figref{alpha}, our model achieves the best PSNR and SSIM scores at $\alpha\!=\!0.125$, which means that 12.5\% of the channels in the real and imaginary parts are reduced to a low spatial frequency. When $\alpha$ becomes larger, the performance quickly degrades due to severe information loss induced by over-large ratios.


The number of network parameters increases as the number of blocks ($bn$) increases. Therefore, it is necessary to choose an appropriate number of blocks to ensure that the network structure reaches the highest reconstruction accuracy without inducing higher computational and memory requirements.
Herein, we carry out various experiments using different numbers of blocks. The results are presented in~\figref{block}.
As can be seen from the curves, our model can successfully reconstruct the MR images at $bn\!=\!4$, and the reconstruction accuracy reaches the highest at $bn\!=\!10$.

Finally, we study the FLOPs of DONet with respect to different $\alpha$ in~\figref{flops}.
The number below each point is the value of $\alpha$, and $\alpha\!=\!0$ refers to the baseline model.
As can be observed, a small $\alpha$ leads to improved performance with a higher FLOPs. Moreover, compared with the baseline model (\eg, $\alpha\!=\!0$), our model consistently shows better performance with much lower FLOPs.



\section{Conclusion}
In this work, we focus on spatial frequency feature expression in complex-valued data for parallel MR image reconstruction. For this purpose, we propose a DONet with a series of novel Dual-OctConv operations to deal with the real and imaginary components of the data at multiple spatial frequencies. 
By convolving the feature maps of both the real and imaginary components under different spatial resolutions,
the proposed Dual-OctConv facilitates our DONet to learn more comprehensive feature representations, yielding higher-quality reconstructed images with significantly reduced artifacts.
We conduct extensive experiments on the $in~vivo$ knee and fastMRI datasets under different settings of undersampling patterns and acceleration rates. The results demonstrate the advantages of our model against state-of-the-art methods in accelerated MR image reconstruction.


\bibliographystyle{IEEEtran}
\bibliography{ref.bib}

\vspace{-8mm}
\begin{IEEEbiography}[{\includegraphics[width=1in,height=1.25in,clip,keepaspectratio]{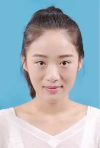}}]{Chun-Mei Feng}
	\small received her M.S. degree at QuFu Normal University, China, in 2018. She is currently a Ph.D. student of the school of Information Science and Technology, Harbin Institute of Technology Shenzhen, China. Her research interests include medical imaging and bioinformatics.
\end{IEEEbiography}
\vspace{-8mm}
\begin{IEEEbiography}[{\includegraphics[width=1in,height=1.25in,clip,keepaspectratio]{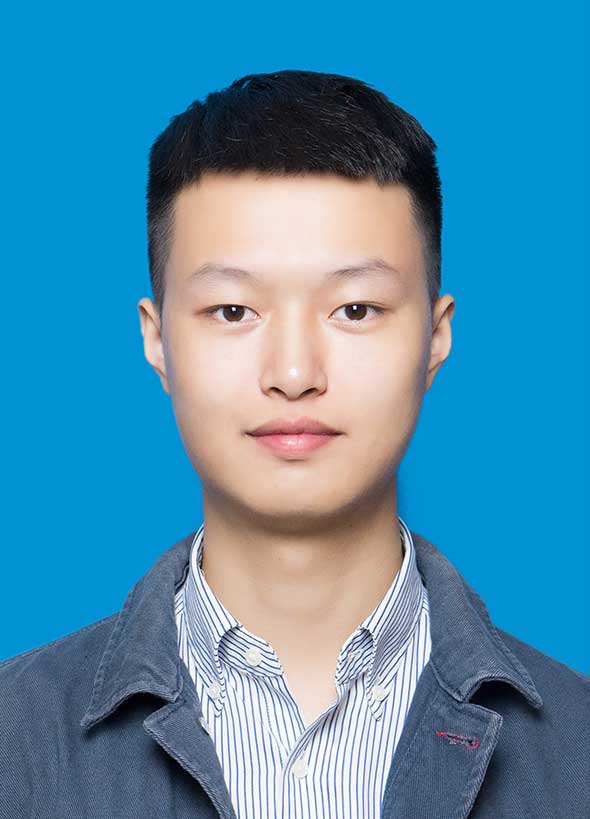}}]{Zhanyuan Yang}
	\small received his M.S. degree at University of Electronic Science and Technology of China (UESTC), China, in 2020. His research interests are computer vision based on deep learning, including image restoration and image enhancement.
\end{IEEEbiography}
\vspace{-8mm}
\begin{IEEEbiography}[{\includegraphics[width=1in,height=1.25in,clip,keepaspectratio]{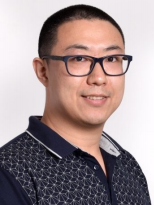}}]{Huazhu Fu}
	\small (SM'18) is a Senior Scientist at Inception Institute of Artificial Intelligence, Abu Dhabi, United Arab Emirates. He received his Ph.D. from Tianjin University in 2013, and was a Research Fellow at Nanyang Technological University for two years. From 2015 to 2018, he was a Research Scientist in Institute for Infocomm Research at Agency for Science, Technology and Research. His research interests include computer vision, machine learning, and medical image analysis. He currently serves as an Associate Editor of IEEE Transactions on Medical Imaging, IEEE Journal of Biomedical and Health Informatics, and IEEE Access. He also serves as a co-chair of OMIA workshop and co-organizer of ocular image series challenge (i-Challenge). 
\end{IEEEbiography}
\vspace{-8mm}
\begin{IEEEbiography}[{\includegraphics[width=1in,height=1.25in,clip,keepaspectratio]{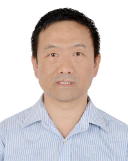}}]{Yong Xu}
	\small (Senior Member, IEEE) received his B.S. and M.S. degrees at Air Force Institute of Meteorology (China) in 1994 and1997, respectively. He then received his Ph.D. degree in pattern recognition and intelligence system at the Nanjing University of Science and Technology in 2005. From May 2005 to April 2007, he worked at Harbin Institute of Technology Shenzhen as a post-doctoral research fellow. Now he is a professor at HIT Shenzhen. His current interests include pattern recognition, machine learning, and bioinformatics.
\end{IEEEbiography}
\vspace{-8mm}
\begin{IEEEbiography}[{\includegraphics[width=1in,height=1.25in,clip,keepaspectratio]{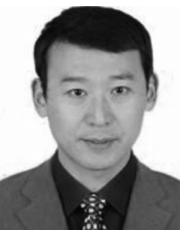}}]{Jian Yang}
	\small (Member, IEEE) received the PhD degree from Nanjing University of Science and Technology (NUST), on the subject of pattern recognition and intelligence systems in 2002. In 2003, he was a Postdoctoral researcher at the University of Zaragoza. From 2004 to 2006, he was a Postdoctoral Fellow at Biometrics Centre of Hong Kong Polytechnic University. From 2006 to 2007, he was a Postdoctoral Fellow at Department of Computer Science of New Jersey Institute of Technology. Now, he is a Chang-Jiang professor in the School of Computer Science and Technology of NUST. He is the author of more than 200 scientific papers in pattern recognition and computer vision. His papers have been cited more than 6000 times in the Web of Science, and 18000 times in the Scholar Google. His research interests include pattern recognition, computer vision and machine learning. Currently, he is/was an associate editor of Pattern Recognition, Pattern Recognition Letters, IEEE Trans. Neural Networks and Learning Systems, and Neurocomputing. He is a Fellow of IAPR.
\end{IEEEbiography}
\vspace{-8mm}
\begin{IEEEbiography}[{\includegraphics[width=1in,height=1.25in,clip,keepaspectratio]{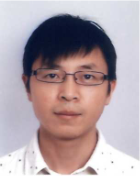}}]{Ling Shao}
	\small (Fellow, IEEE) is the CEO and Chief Scientist of the Inception Institute of Artificial Intelligence, Abu Dhabi, United Arab Emirates. He is an Associate Editor of the IEEE TRANSACTIONS ON IMAGE PROCESSING, the IEEE TRANSACTIONS ON NEURAL NETWORKS AND LEARNING SYSTEMS, the IEEE TRANSACTIONS ON CIRCUITS ANDSYSTEM FOR VIDEO TECHNOLOGY, and several other journals. His research interests include Computer Vision, Machine Learning and Medical Imaging. He is a Fellow of the IEEE, the IAPR, the IET and the BCS.
\end{IEEEbiography}
\end{document}